\documentclass[aps,prb,twocolumn,groupedaddress]{revtex4-1}
\usepackage{graphicx} 
\usepackage[dvipdfm]{hyperref}
\hypersetup{colorlinks=true,linkcolor=blue,
  implicit=true,breaklinks=true,pagebackref=true,backref=true,
bookmarks=true,bookmarksnumbered=true,hyperfootnotes=true,debug=true,
  naturalnames=false,citecolor=blue,pdfview=FitH,pdfstartview=FitH,hyperindex=true}
\usepackage{amsmath}
\usepackage{amssymb}
\usepackage{ifsym}
\begin{document}

\title{Pressure-tuned superconductivity and normal-state behavior in Ba(Fe$_{0.943}$Co$_{0.057}$)$_2$As$_2$ near the antiferromagnetic boundary}
\author{W. Liu$^{1,2}$}
\author{Y. F. Wu$^{3,4}$}
\author{X. J. Li$^{1}$}
\author{S. L. Bud$^{'}$ko$^{5}$}
\author{P. C. Canfield$^{5}$}
\author{C. Panagopoulos$^{6}$}
\author{P. G. Li$^{2}$}
\author{G. Mu$^{3}$}
\author{T. Hu$^{3\dag}$}
\author{C. C. Almasan$^{^7}$}
\author{H. Xiao$^{1*}$}

\affiliation{$^{1}$ Center for High Pressure Science and Technology Advanced Research, Beijing, 100094, China}
\affiliation{$^{2}$ Department of Physics, Zhejiang SCI-TECH University, Hangzhou, 310018 China}
\affiliation{$^{3}$ State Key Laboratory of Functional Materials for Informatics, Shanghai Institute of Microsystem and Information Technology, Chinese Academy of Sciences, 865 Changning Road, Shanghai 200050, China}
\affiliation{$^{4}$ University of Chinese Academy of Science, Beijing 100049, China}
\affiliation{$^{5}$ Ames Laboratory and Department of Physics and Astronomy, Iowa State University, Ames, Iowa 50011}
\affiliation{$^{6}$ Division of Physics and Applied Physics, School of Physical and Mathematical Sciences, Nanyang Technological University, 637371 Singapore}
\affiliation{$^{7}$ Department of Physics, Kent State University, Kent, Ohio, 44242, USA}

\date{\today}
\begin{abstract}
Superconductivity in iron pnictides is unconventional and pairing may be mediated by magnetic fluctuations in the Fe-sublattice. Pressure is a clean method to explore superconductivity in iron based superconductors by tuning the ground state continuously without introducing disorder.  Here we present a systematic high pressure transport study in Ba(Fe$_{1-x}$Co$_{x}$)$_2$As$_2$ single crystals with $x=$ 0.057, which is near the antiferromagnetic instability. Resistivity $\rho=\rho_0+AT^{n}$ was studied under applied pressure up to 7.90 GPa. The parameter $n$ approaches a minimum value of $n\approx 1$ at a critical pressure $P_c =$ 3.65 GPa. Near $P_c$, the superconducting transition temperature $T_c$ reaches a maximum value of 25.8 K. In addition, the superconducting diamagnetism at 2 K shows a sudden change around the same critical pressure.
These results may be associated with a possible quantum critical point hidden inside the superconducting dome, near optimum $T_c$.
\end{abstract}

\pacs{ 74.25.Dw  74.25.Uv  74.40.Kb  74.72.Kf  }

\maketitle
\subsection{Introduction}
Unconventional superconductivity observed in iron-based superconductors is in close proximity to an antiferromagnetically ordered state.\cite{Stewart2011} Superconductivity emerges as antiferromagnetism is suppressed by pressure or chemical doping,\cite{Ni2008, Colombier2009, Canfield2010} and the superconducting critical temperature $T_c$ forms a dome shape. In the Ni-, Co-, P-, Rh- and Pd-doped BaFe$_2$As$_2$ system, the antiferromagnetic phase boundary crosses the superconducting dome near optimal doping.\cite{Ni2008, Ni2010, Luo2012, Hashimoto2012, Canfield2009, Ni2009, Ni2008, Chu2009} Hence, there is a region in the phase diagram where antiferromagnetism and superconductivity coexist.
Neutron scattering measurements on Ba(Fe$_{1-x}$Ni$_x$)$_2$As$_2$ observed short range incommensurate antiferromagnetic order coexisting with superconductivity near optimal doping, where the first-order-like antiferromagnetism-to-superconductivity transition suggests the absence of a quantum critical point (QCP).\cite{Luo2012} Notably, it has been reported that the magnetic penetration depth in BaFe$_2$(As$_{1-x}$P$_x$)$_2$ shows a sharp peak at optimal doping, possibly due to quantum fluctuations associated with a QCP.\cite{Hashimoto2012}

In particular for Ba(Fe$_{1-x}$Co$_x$)$_2$As$_2$, the physical properties have been widely studied close to optimal doping and the antiferromagnetic phase boundary. Neutron diffraction measurements indicate Co doping rapidly suppresses antiferromagnetism, with the antiferromagnetic order vanishing at $x \approx$ 0.055.\cite{Lester2009} For $x=$ 0.06, it is suggested that superconductivity coexists with a spin density wave (SDW).\cite{Huang2017a} For thin films of Ba(Fe$_{1-x}$Co$_{x}$)$_2$As$_2$, the exponent $n$ in the temperature dependence of the resistivity is minimum namely, close to unity at $x\approx$ 0.05 and $x\approx$ 0.07 for MgO and CaF$_2$ substrate, respectively, which may be associated with an antiferromagnetic QCP. \cite{Iida2016} Furthermore, a sign change in the electronic-magnetic Gruneisen parameter is observed for $x=0.055$ and $x=0.065$, consistent with the expected behavior at a QCP.\cite{Meingast2012} In addition, a critical concentration of $x_c \approx 0.065$ is determined from the analysis of $1/T_{1}T$ in NMR measurements. \cite{Ning2014} Considerably enhanced flux-flow resistivity $\rho_{ff}$ was also detected for $x=$ 0.06, perhaps due to enhancement of spin fluctuations near QCP.\cite{Huang2017}  Thermopower($S$) measurements reported a maximum $S/T$ in proximity to the commensurate-to-incommensurate SDW transition for $x \approx$ 0.05, close to the highest superconducting $T_c$.\cite{Arsenijevifmmodecuteclseci2013} However, the superconducting magnetization appears nearly unchanged across the dome in Ba(Fe$_{1-x}$Co$_x$)$_2$As$_2$.\cite{Ni2008}

Despite extensive studies in Ba(Fe$_{1-x}$Co$_x$)$_2$As$_2$ close to optimal doping, there had been no systematic study on how the normal state evolves across the antiferromagnetic phase boundary. Here we probe the phase diagram close to the antiferromagnetic boundary through measurements of resistivity and magnetization by tuning the applied pressure in a sample with $x=$ 0.057. Normal state resistivity changes from non-Fermi liquid to Fermi liquid with increasing  pressure. It shows almost linear temperature dependence at a critical pressure of $P=$ 3.65 GPa, where $T_c$ is maximum. In addition, the residual resistivity $\rho_0$ and the resistivity at $T_c$ all change around the same critical pressure.
From the magnetization data, the superconducting diamagnetism at 2 K shows a sudden change at a critical pressure of $P=$ 3.5 GPa, in accordance with changes in resistivity.
These results may be due to a possible QCP at optimum $T_c$, similar to the case of BaFe$_2$(As$_{1-x}$P$_x$)$_2$ \cite{Hashimoto2012} and hole doped cuprates.\cite{Sachdev2010}


\subsection{Experimental Details}
Single crystals of Ba(Fe$_{1-x}$Co$_{x}$)$_2$As$_2$ with $x=0.057$ were synthesized by a flux method.\cite{Ni2008}
Electrical resistivity was measured using a Quantum Design Physical Property Measurement System (PPMS). The electronic transport properties were measured using four-probe electrical conductivity in a diamond anvil cell made of CuBe alloy. The diamond culet was 800 $\mu$m in diameter.  Magnetic measurements were performed in a superconducting quantum interference device (SQUID magnetometer). Pressure was applied using a diamond anvil cell made of CuBe alloy with the diamond anvil culet of 500 $\mu$m. In both cases, Daphne oil 7373 was used as a pressure-transmitting medium. Above its solidification at 2.2 GPa,\cite{Sefat2011} non-hydrostaticity  may develop and lead to inhomogeneous pressure distribution inside the sample chamber. Pressure was calibrated by using the ruby fluorescence shift at room temperature. For resistivity, the superconducting transition temperature $T_c$ is defined as the temperature for the appearance of zero resistance state (Fig. 1(b)); for magnetization, $T_c$ is the temperature we observe a sharp drop in $M$ (inset to Fig. 3(a)).

\subsection{Results and Discussion}
Figure 1 shows the temperature  dependence of resistivity for Ba(Fe$_{1-x}$Co$_{x}$)$_2$As$_2$ with $x=$ 0.057 measured at different applied pressures namely, $P=$ 0, 1.25, 2.69, 3.65, 5.26, 6.87 and 7.90 GPa. The resistivity curve for $P=$ 7.90 GPa was shifted downward by 0.05 m$\Omega$ cm for clarity. Note that the large decrease of $\rho_{300K}$ with pressure (inset to Fig. 1(b)) is very similar to the changes occurring with Co doping in Ba(Fe$_{1-x}$Co$_{x}$)$_2$As$_2$.\cite{Ahilan2009}
By comparing the data we find that an increase in doping level by 1$\%$ is roughly equivalent to 1.2 GPa of pressure, which is comparable with previous report.\cite{Meingast2012}

At low pressures, resistivity decreases with decreasing temperature but shows an upturn just before entering the superconducting state. This upturn is due to the structural ($T_s$) and SDW ($T_{sdw}$) phase transition, in agreement with earlier studies in underdoped  Ba(Fe$_{1-x}$Co$_{x}$)$_2$As$_2$.\cite{Ni2008} Both $T_s$/$T_{sdw}$ can be estimated from the first derivative of the temperature dependent resistivity curve (see inset to Fig. 1(a)).\cite{Ni2010} With further increase in pressure, the upturn vanishes suggesting suppression of the $T_s$ and $T_{sdw}$. Similar changes with pressure has been reported for Ba(Fe$_{1-x}$Co$_{x}$)$_2$As$_2$.\cite{Ahilan2008, Ahilan2009, Colombier2010} The zero resistance transition temperature $T_c$ (solid squares in Fig. 2(a)) varies non-monotonically with increasing pressure. For $P=$ 6.87 GPa and above, we observe a finite resistivity down to the lowest measured temperature. A similar dome shaped variation in $T_c$ is observed in Ba(Fe$_{1-x}$Co$_{x}$)$_2$As$_2$ with Co doping.\cite{Colombier2010, Ahilan2008}

\begin{figure}
\centering
\includegraphics[trim=0cm 0cm 0cm 0cm, clip=true, width=0.45\textwidth]{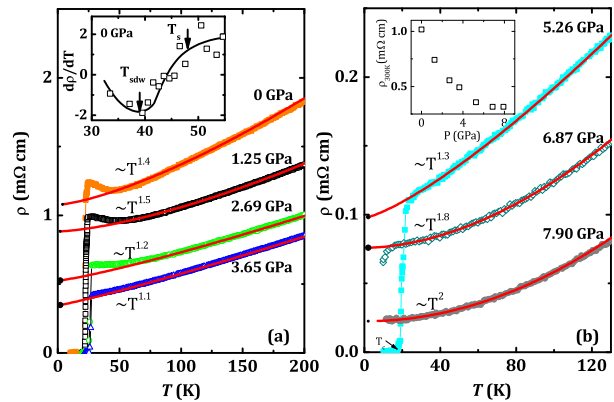}
\caption{\label{fig:magnetoresistance} (a)(b) Temperature $T$ dependence of resistivity $\rho$ under applied pressure $P=$ 1.25, 2.69, 3.65, 5.26, 6.87 and 7.90 GPa. Symbols represent data and solid lines are fits using $\rho=\rho_0+AT^{n}$. Note that the resistivity curve for $P=$ 7.90 GPa was shifted downward by 0.05 m$\Omega$ cm for clarity. Inset to (a) shows  the temperature dependence of d$\rho$/d$T$  at ambient pressure. Inset to (b) shows pressure dependence of resistivity at 300 K, $\rho_{300K}$.}
\end{figure}

We fit the resistivity curve under pressure using $\rho=\rho_0+AT^{n}$ (with fitting parameters $\rho_0$, $n$ and $A$) as shown in Fig. 1, where the symbols represent data points and the solid lines are fits.
The pressure dependence of $T_c$, $\rho_0$, $\rho$ at $T_c$ and $n$  obtained from Fig. 1 are summarized in Fig. 2(a)-(c), respectively. Resistivity can be tuned with pressure from a non-Fermi liquid (NFL)($n=$ 1) to Fermi liquid (FL) ($n=$ 2) behavior. Note that $n=1.1$ at $P=$ 3.65 GPa and increases with further increase in pressure, reaching 2 at $P=$ 7.90 GPa.

Interestingly, all parameters in Fig. 2 show a change at $P_c \approx $ 3.5 GPa. This is similar to the heavy fermion superconductor CeCoIn$_5$, where $\rho_0$ and $n$  change at $P_c=$ 1.6 GPa.\cite{Sidorov2002}
We ascribe the decrease in $\rho_0$ with increasing pressure to
a change in inelastic scattering.\cite{Sidorov2002} The pressure dependence of $\rho$ at ${T_c}$ shows a change in slope at $P_c$, similar to the behavior of the normal state resistivity $\rho_n$ at $T_c$ around optimal doping in chemically tuned BaFe$_2$As$_2$.\cite{Huang2017a}
Similar change in $n$ was also observed in BaFe$_2$As$_2$ with Co doping, where the exponent $n$ is minimum namely, close to 1 at optimal doping.\cite{Iida2016} In BaFe$_2$(As$_{1-x}$P$_{x})_2$, non-Fermi liquid behavior with $n$ close to unity is found around optimal doping $x=$ 0.3, with $T_c$ maximum at the QCP.\cite{Hashimoto2012} Similarly, linear resistivity was observed for Ba(Fe$_{1-x}$Ni$_{x}$)$_2$As$_2$ with $x=$ 0.05 for which $T_c$ is maximum at a magnetic QCP.\cite{Zhou2013}

\begin{figure}
\centering
\includegraphics[trim=0cm 0cm 0cm 0cm, clip=true, width=0.45\textwidth]{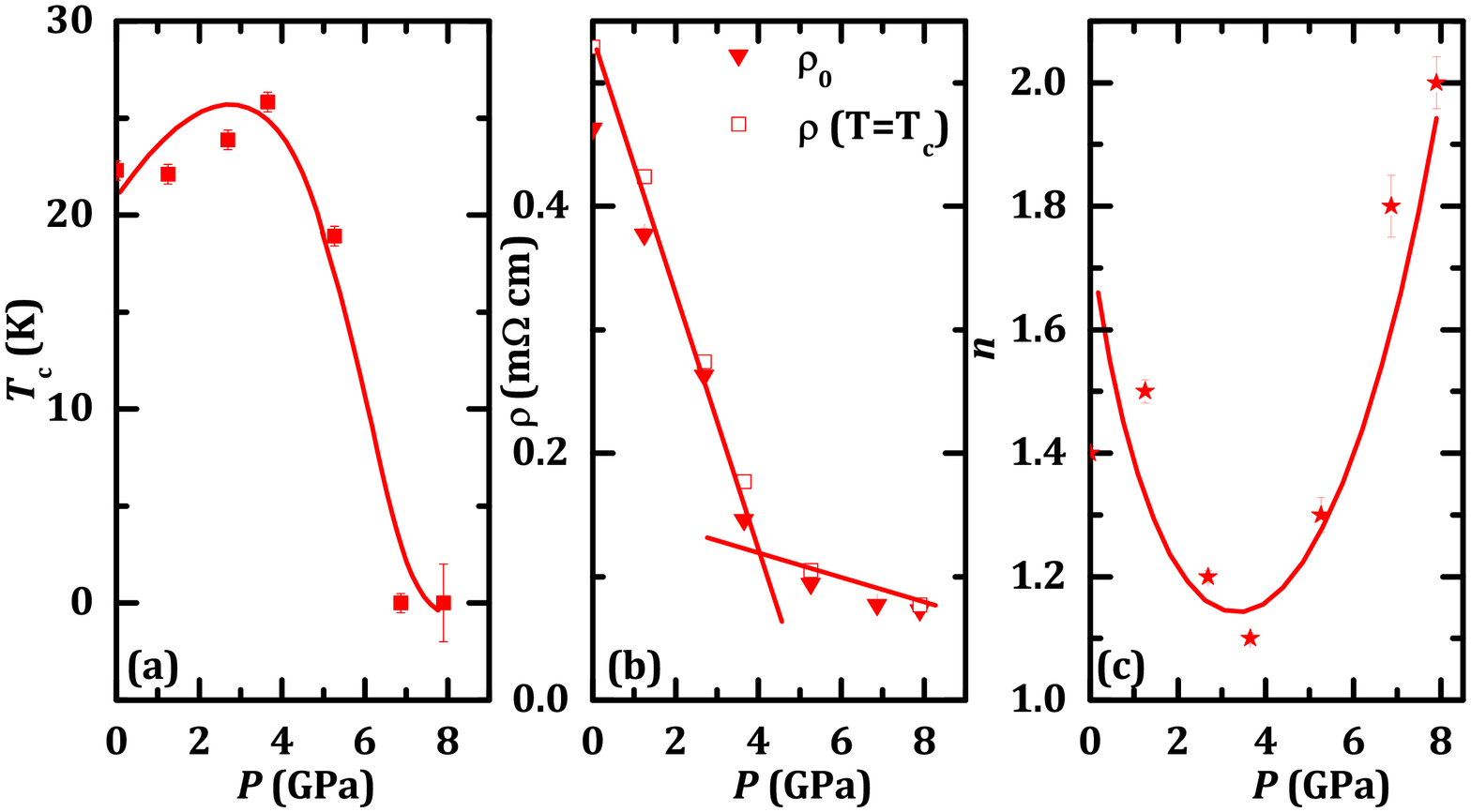}
\caption{\label{fig:magnetoresistance}  Pressure dependence of (a) superconducting transition temperature $T_c$, (b) resistivity at the superconducting onset temperature $\rho(T=T_c)$ and residual resistivity $\rho_0$, (c) exponent $n$.
}
\end{figure}

The zero field cooled (ZFC) magnetization was measured in a run with increasing pressure for $P=$  0.6, 1.2, 2.0, 2.7, 3.5, 4.3, 5.6, 6.4 GPa. The resultant data are plotted in Fig. 3(a). Since the sample used in the pressure cell is too small to measure its mass, we show magnetization data in emu. Another piece of sample is used to obtain the ambient pressure magnetization data (as shown in the inset to Fig. 3(a)) to determine $T_c$ at $P=0$.  The pressure dependence of $T_c$ determined from magnetization measurements is plotted in Fig. 3(b), consistent with the $T_c$ obtained from resistivity measurements (Fig. 2(a)).

\begin{figure}
\centering
\includegraphics[trim=0cm 0cm 0cm 0cm, clip=true, width=0.45\textwidth]{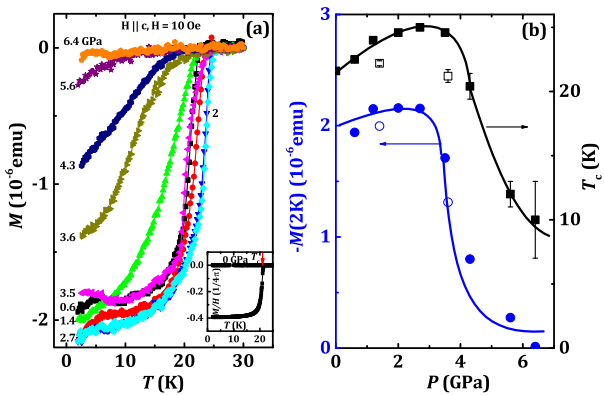}
\caption{\label{fig:magnetoresistance} (a) Temperature dependence of magnetization measured at $P=$ 0.6, 1.2,  2, 2.7, 3.5, 4.3, 5.6, 6.4 GPa with increasing pressure and  3.6, 1.4 GPa with decreasing pressure, in an applied magnetic field of 10 Oe. The inset shows magnetization data at ambient pressure for both zero field cooled (ZFC) and field cooled (FC) runs. (b) Pressure dependence of the superconducting transition temperature (squares) and the diamagnetic signal $M$(2K) (circles). Solid and open symbols depict data for experiments performed with increasing and decreasing pressure, respectively. }
\end{figure}

We summarize the pressure dependence of the ZFC magnetization at $T=$ 2 K, $M$(2K) in Fig. 3(b). Note that the magnetization data at low temperatures was often used to estimate the superconducting volume fraction.\cite{Saha2009a, Saha2012, Kudo2013} In our case, it may not be accurate to estimate the volume fraction of superconductivity from magnetization since the superconducting transitions are broad and incomplete at high pressures and upon releasing the pressure. Nevertheless, it will give some hint to further understand the behavior of the superconducting state evolving across the antiferromagnetic phase boundary.
Initially, $M_{2K}$ slightly increases with pressure followed by a sudden suppression at $P_c=$ 3.5 GPa, then becoming negligible at high pressures.
A similar pressure induced suppression in the superconducting volume was observed in the parent compound of BaFe$_2$As$_2$ and SrFe$_2$As$_2$, where a dome like behavior of the pressure dependent superconducting volume is reported.\cite{Alireza2009} Also, for Sr(Fe$_{1-x}$Ni$_{x}$)$_2$As$_2$ and Ca$_{1-x}$LaFe$_2$(As$_{1-y}$P$_{y}$)$_2$, the superconducting volume shows a dome behavior with doping.\cite{Saha2009a, Kudo2013}  In addition, a sudden suppression in the superconducting volume was observed in high-$T_{c}$ cuprate La$_{2-x}$Sr$_{x}$CuO$_{4}$ at a critical doping level of around $x=$ 0.21,\cite{Takagi1992} which is close to a QCP.\cite{Cooper2009} Thus, the suppression of the superconducting volume fraction above the critical pressure observed in present work could reflect a phase transition at $P_c$.

 Note that in chemically doped (Co, Rh, Ni) BaFe$_2$As$_2$ at ambient pressure, there is no change in magnetization across the dome.\cite{Ni2008, Ni2009, Ni2010} Nevertheless, this difference may be due to different role played by pressure and chemical tuning. In fact, there is a pressure tuned QCP in pure CeCoIn$_5$,\cite{Sidorov2002} while, there is no signatures of quantum critical behavior in Cd-doped CeCoIn$_5$, due to the effect of disorder near a zero temperature magnetic instability.\cite{Seo2013} This suggests that tuning a system with disorder to a presumed magnetic QCP does not necessitate a quantum critical response. \cite{Seo2013}

We also measured two magnetization curves under decompression, namely, for $P=$ 3.6 and 1.4 GPa (see Fig. 3(a)). Interestingly, the superconducting volume fraction is about the same as compression data, however, the $T_c$ values are not fully recovered. The different $T_c$ between compression and decompression is previously reported in In$_2$Se$_3$, which is intrinsic, as a result of changes in phonon and variation of carrier concentration combined in the pressure quench.\cite{Ke2017}
 Further measurements are needed to confirm if there is indeed a suppressed $T_c$ behavior in Ba(Fe$_{1-x}$Co$_{x}$)$_2$As$_2$ during decompression, which is beyond the scope of this work.

\begin{figure}
\centering
\includegraphics[trim=0cm 0cm 0cm 0cm, clip=true, width=0.45\textwidth]{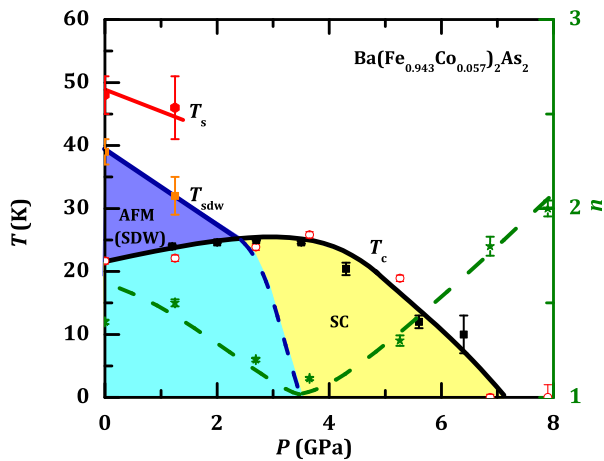}
\caption{\label{fig:magnetoresistance} Temperature - pressure ($T-P$) phase diagram of Ba(Fe$_{1-x}$Co$_{x}$)$_2$As$_2$ with $x=$ 0.057. The structural phase transition temperature $T_s$ is marked as red hexagon. The SDW phase transition temperature $T_{sdw}$ is marked as orange squares.  The superconducting transition temperature $T_c$ is determined from magnetization (solid squares) and  resistivity  (open circles) measurements. The exponent $n$ is indicated by stars.
The light blue and yellow represent the region with large and small superconducting diamagnetism, respectively.
}
\end{figure}

Figure 4 shows the temperature vs. pressure ($T-P$) phase diagram of Ba(Fe$_{1-x}$Co$_{x}$)$_2$As$_2$ with $x=$ 0.057.  The structural phase transition temperature ($T_s$), the SDW antiferromagnetic phase transition temperature $T_{sdw}$, the superconducting transition temperature $T_c$ and the exponent $n$ in $\rho=\rho_0+AT^{n}$ are summarized.
With increasing pressure, we observe a suppression of the antiferromagnetic phase whereas, the superconducting transition temperature increases, suggesting competition between the two. $T_c$ reaches a maximum at a critical pressure $P_c$ around 3.5 GPa and decreases with further increase in pressure, forming a dome shape. Around $P_c$, we observe signature of a non-Fermi liquid namely, $n$ close to 1, often associated with quantum criticality.\cite{Cooper2009, Gegenwart2008}
This is accompanied by the above mentioned change in the superconducting diamagnetism. Together,
these experimental findings suggest the presence of a QCP at $P_c$, where $T_c$ is maximum.

Earlier NMR measurements in Ba(Fe$_{1-x}$Co$_{x}$)$_2$As$_2$ revealed that the maximum $T_c$ occurs at the antiferromagnetic QCP possibly due to magnetically mediated superconductivity.\cite{Nakai2013} Such a superconducting pairing  mechanism may be applicable in several strongly correlated superconducting systems, where fundamental physical quantities, including the superconducting condensation energy, quasiparticle lifetime, and superfluid density show abrupt changes at a QCP.\cite{Panagopoulos2002a} Hence, the observation of a linear temperature dependence of resistivity at $P_c$ about 3.5 GPa and a possible change in the superconducting volume fraction, may be associated with a quantum phase transition.

\subsection{Conclusions}
In summary, electrical resistivity and magnetization under pressure were measured in Ba(Fe$_{1-x}$Co$_{x}$)$_2$As$_2$ with $x=$ 0.057. Resistivity shows linear temperature dependence around a critical pressure of 3.5 GPa where $T_c$ is maximum. Furthermore, we detected signs of an accompanied change in the superconducting volume. These results are most likely due to a possible pressure tuned QCP hidden inside the superconducting dome of Ba(Fe$_{1-x}$Co$_{x}$)$_2$As$_2$.

\subsection{Acknowledgments}
 Work at HPSTAR was supported by NSAF, Grant No. U1530402. Work at SIMIT was supported by the support of NSFC, Grant No. 11574338. Work at the Ames Laboratory was supported by the U.S. Department of Energy, Basic Energy Sciences under Contract No. DE-AC02-07CH11358. The work in Singapore was supported by the National Research Foundation ¨C NRF Investigatorship (Reference No. NRF-NRFI2015-04). The work at KSU was supported by the National Science Foundation under grant  No. DMR-1505826.

$^{*}$ hong.xiao@hpstar.ac.cn $^{\dag}$ thu@mail.sim.ac.cn

W.L. and Y.F.W. contributed equally to this work.
%

\end{document}